\documentclass[a4paper,11pt]{article}
\pdfoutput=1 

\usepackage{jcappub} 

\usepackage[T1]{fontenc} 

\title{\boldmath Measuring the transverse baryonic acoustic scale from the SDSS DR11 galaxies}

\author[a,1]{G. C. Carvalho,\note{Corresponding author.}}
\author[b]{A. Bernui,}
\author[c,d]{M. Benetti}
\author[b]{J. C. Carvalho}
\author[b,e]{E. de Carvalho}
\author[b]{J. S. Alcaniz}

\affiliation[a]{ Universidade do Estado do Rio de Janeiro, Faculdade de Tecnologia, 27537-000, Resende - RJ, Brasil}
\affiliation[b]{ Observat\'orio Nacional, 20921-400, Rio de Janeiro - RJ, Brasil}
\affiliation[c]{ Dipartimento di Fisica "E. Pancini", Universit\`a di Napoli "Federico II", Via Cinthia, I-80126, Napoli, Italy}
\affiliation[d]{ Istituto Nazionale di Fisica Nucleare (INFN), sez. di Napoli, Via Cinthia 9, I-80126 Napoli, Italy}
\affiliation[e]{ Centro de Estudos Superiores de Tabatinga, Universidade do Estado do Amazonas, 69640-000, Tabatinga - AM, Brasil }

\emailAdd{gabriela.coutinho@fat.uerj.br}
\emailAdd{bernui@on.br}
\emailAdd{benettim@na.infn.it}
\emailAdd{jcarvalho@on.br}
\emailAdd{edfilho@uea.edu.br}
\emailAdd{alcaniz@on.br}

\abstract{We report five measurements of the transverse baryonic acoustic scale, $\theta_{BAO}$, obtained from the angular two-point correlation 
function calculation for Luminous Red Galaxies of the eleventh data release of the Sloan Digital Sky Survey (SDSS). Each measurement 
has been obtained by considering a thin redshift shell ($\delta z = 0.01$ and $0.02$) in the interval $ z \in [0.565, 0.660] $, which 
contains a large density of galaxies ($\sim 20,000$ galaxies/redshift shell). Differently from the three-dimensional Baryon Acoustic Oscillations (BAO) measurements, 
these data points are obtained almost model-independently and provide a Cosmic Microwave Background (CMB)-independent way to estimate the sound horizon $ r_s $. 
Assuming a time-dependent equation-of-state parameter for the dark energy, we also discuss  constraints on the main cosmological 
parameters from $\theta_{BAO}$ and CMB data.}

\begin{document}
\maketitle
\flushbottom

\section{Introduction}
\label{sec1}
The Baryon Acoustic Oscillations (BAO)  left mensurable signatures in the distribution of galaxies which have been robustly detected from data of galaxy redshift surveys~\cite{Eisenstein05,Cole05,Percival10,Beutler11,Blake11b,Sanchez12,Salazar16}. In particular, the Sloan Digital Sky Survey (SDSS) has released, over sixteen years~\cite{SDSS}, increasing 3-dimensional galaxy catalogs, making possible precise measurements of the BAO scale at various redshifts. Such  measurements, along with data from other cosmological observables, e.g., type Ia Supernovae (SNIa)~\cite{sne} and the Cosmic Microwave Background (CMB)~\cite{wmap9,planck}, provide currently the most  precise constraints on the late-time evolution of the Universe~\cite{Peebles17}. 

As is well known, the BAO signature defines a statistical standard ruler and provides independent estimates of the the Hubble parameter $H(z)$ and the angular diameter distance $D_{\!A}(z)$ through the radial ($dr_{\parallel} = c\delta z/H(z)$) and transverse ($dr_{\perp} = (1+z)D_A\theta_{BAO}$) BAO modes, respectively (for a recent review, see \cite{eisensteinreview}).  Measurements of the BAO scale are usually obtained from the application of the spatial 2-point  correlation function (2PCF) to a large distribution of galaxies, where the BAO signature appears as a bump at the corresponding scale. Analysis of this type  assumes a fiducial cosmology in order  to transform the measured angular positions and redshifts  into  comoving distances and, as discussed in Refs. \cite{Eisenstein05, Sanchez12} (see also \cite{Salazar16}), such conversion  may  bias  the  parameter  constraints.  

Another possibility is to use the angular 2-point correlation function (2PACF), $w(\theta)$, which  involves only the angular separation $\theta$ between pairs, yielding information of $D_{A}(z)$ almost model-independently, provided that the comoving sound horizon $r_s$ is known.  In order to extract useful information from the 2PACF,  the galaxy sample is divided into redshift shells whose width must be quite narrow ($\delta z \sim 10^{-2}$) to avoid large projection effects from the radial BAO signal~\cite{Sanchez11,Carnero,Salazar16,Carvalho}. In a recent analysis, Carnero et al.~\cite{Carnero} provided the first determination of the angular BAO scale $\theta_{\rm{BAO}}$ using $\simeq 0.6 \times 10^6$ Luminous Red Galaxies (LRGs), selected from the SDSS imaging data, with photometric redshifts. Covering the redshift interval $z \in [0.5,0.6]$ (with bin size of $\delta z = 0.1$), they obtained $\theta_{\rm{BAO}} (\bar{z}=0.55) = (3.90 \pm 0.38)^{\circ}$. More recently, Carvalho et al.~\cite{Carvalho} used a different methodology to measure $\theta_{\rm{BAO}}$ from the LRGs sample of the tenth data release of the  SDSS. In that work, six new measurements of $\theta_{\rm{BAO}}$ were obtained from a sample of 409,337 LRGs divided into narrow redshift bins of size $\delta z = 0.01$ and 0.02, with mean redshifts ranging in the interval $\bar{z} = 0.45 - 0.55$. This latter analysis was complemented by two other measurements of the transverse BAO scale at $\bar{z} = 0.235$ and $\bar{z} = 0.365$ obtained from 105,831 LRGs from the seventh data release of the SDSS~\cite{Alcaniz} and at $\bar{z} = 2.225$ using 10,526 quasars from twelfth public Data Release Quasar catalogue (DR12Q) \cite{deCarvalho}.

The goal of the present work is to extend the previous analyses in the following directions. First, we use a denser LRGs catalog, i.e., the eleventh data release of the SDSS (DR11), which includes 543,116 LRGs lying in the redshift interval $z \in [0.43,0.70]$. Second, we obtain the BAO angular scale at higher redshifts, which turns out to improve our final constraints on the cosmological parameters. Finally, from this distribution of LRGs, we also provide a determination of the sound horizon $r_s$ whose uncertainty is comparable with those estimated by current CMB data.

This work is organized as follows. In Sec. \ref{sec2} we explain the methodology used to measure the BAO angular scale $\theta_{BAO}$ using 2PACF. The data set used in the analysis is described in Sec. \ref{sec3}.  The new five measurements of the BAO angular scale are presented and discussed in Sec. \ref{sec4}.  In Sec. \ref{sec5}, assuming a time-dependent parameterization of the dark energy equation of state, $w(z)$, we investigate the constraints on the dark energy  parameters from the $\theta_{\rm{BAO}}$ measurements. We also discuss a CMB-independent estimate of the sound horizon $r_s$. We summarize our main results in Sec. \ref{sec6}.

\section{Methodology} 
\label{sec2}

In what follows, we will describe the methodology used to measure the BAO angular scale $\theta_{\rm{BAO}}$ from the LGRs sample of the SDSS DR11. In order to avoid bias on the cosmological constraints  discussed in Sec. \ref{sec5}, we use an approach as model-independent as possible based on the application of the 2-point angular correlation function (2PACF) considering narrows redshifts bins.  We follow closely the methodology introduced in Ref.~\cite{Carvalho}.

The 2PACF is defined as the excess joint probability that two point sources are found in two solid angle elements $d\Omega_1$ and $d\Omega_2$ with angular separation $\theta$ compared to a homogeneous Poisson distribution~\cite{Peebles-Hauser} and is obtained by comparing the real catalog with random catalogs that follow the geometry of the survey. Here, we adopt the Landy-Szalay estimator~\cite{Landy-Szalay}, 
\begin{equation} \label{ls}
w(\theta) \,=\, \frac{DD(\theta) \,-\, 2 DR(\theta) \,+\, RR(\theta)}{RR(\theta)} \, , 
\end{equation}
where $DD(\theta)$ and $RR(\theta)$ correspond to the number of galaxy pairs with angular separation $\theta$ in 
data-data and random-random catalogs, respectively, whereas $DR(\theta)$ corresponds to the number 
of pairs with separation $\theta$ calculated between a data-galaxy and a random-galaxy. The transverse signal of the acoustic BAO scale
manifests itself as a bump at certain angular scale $\theta_{\rm{FIT}}$, which does not takes into account the corrections from the projection effects discussed in Sec. \ref{sec4}.

Theoretically, the predicted 2PACF, $w_{E}$, can be obtained from the 2PCF, $\xi_{E}$, as 
%
\begin{equation} \label{expected}
w_{E}(\theta, \bar{z}) = \int_0^\infty dz_1 \,\phi(z_1) \int_0^\infty dz_2 \,\phi(z_2) \,\xi_{E}(s, \bar{z}) \, , 
\end{equation}
where $\bar{z} \equiv (z_1 + z_2) / 2$, with $z_2 = z_1 + \delta z$,   $\phi(z_i)$ is the normalised galaxy selection function at $z_i$ and 
\begin{equation}
\xi_{E}(s,z)=\! \int_0^\infty \! \frac{dk}{2\pi^2} \, k^2 \, j_0(k s) \, b^2 \, P_m(k, z) \, .
\label{xi_e}
\end{equation}
In the above expression, $j_0$ is the zeroth order Bessel function, $b$ is the galaxy bias and $P_m(k, z)$ is the matter power spectrum. The comoving distance between a pair of galaxies at two redshifts $z_1$ and $z_2$, assuming a flat Friedmann-Lama\^itre-Robertson-Walker geometry, is given by 
\begin{equation}
s = \sqrt{r^2(z_1) \,+\, r^2(z_2) \,-\, 2 \,r(z_1) \,r(z_2)  \cos \theta_{12} \,\,} \, ,
\end{equation}
where $\theta_{12}$ is the angular distance between such pair of galaxies, and the radial distance between the observer and a galaxy at redshift $z_i$, 
\begin{equation} \label{rz}
r(z_i) = c\int_{0}^{z_i}{dz/H(z, \bf{p})}\;, 
\end{equation}
depends on the  parameters ${\bf{p}}$  of the fiducial cosmology adopted in the analysis through the Hubble parameter $H$. Because the bin shells considered in this analysis are narrow, $\delta z \sim 10^{-2}$, then $z_1 \approx z_2$ and $\xi_{E}(s, z_1) \simeq \xi_{E}(s, z_2)$. This amounts to saying that one can safely assume that $\xi_{E}(s, \bar{z})$ depends only on the constant $\bar{z}$ (we refer the reader to~\cite{Carvalho} for a more detailed discussion on this issue).

\begin{table}
\centering
 \begin{tabular}{|c|c|c|}
\hline
$\bar{z}$& $z$ interval &   $N_g$ \\
\hline
\hline
\quad 0.57 \quad &  \quad 0.565 - 0.575  \quad & \quad 24,967 \quad \\
\quad 0.59 \quad &  \quad 0.585 - 0.595  \quad & \quad 21,292 \quad \\
\quad 0.61 \quad &  \quad 0.605 - 0.615  \quad & \quad 18,003 \quad \\
\quad 0.63 \quad &  \quad 0.625 - 0.635  \quad & \quad 14,275 \quad \\
\quad 0.65 \quad &  \quad0.640 - 0.660  \quad & \quad 21,949 \quad \\
\hline 
\end{tabular}
\caption{The five redshift bins considered in the analysis and their numbers: mean redshift, $\bar{z}$, redshift interval and number of galaxies, $N_g$. The contiguous intervals are separated by a redshift interval of size 0.005 to avoid correlation between the bins.}
\label{table1}
\end{table}

\begin{figure*} 
\label{figa}
\begin{center}
\mbox{\hspace{-0.2cm}
\includegraphics[scale=0.65]{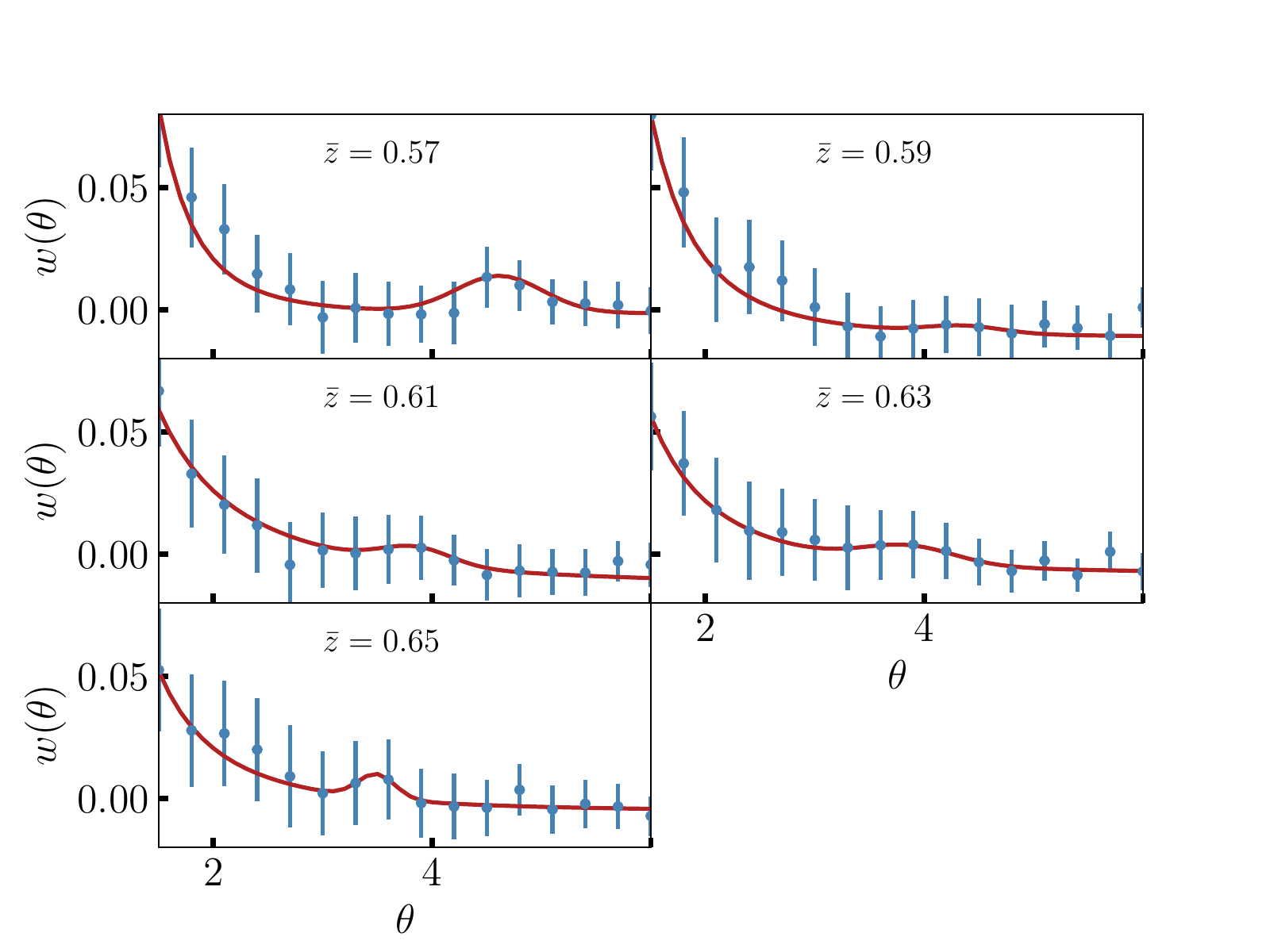}
}
\end{center}
\caption{The 2PACF as a funtion of $\theta$ [deg] for five bin redshift intervals using the DR11-SDSS data (bullets) and Eq. (\ref{eq:fit}) (continuous line). 
The BAO location and the width are related 
to $\theta_{FIT}$ and $\sigma_{FIT}$, respectively.}
\label{fig1}
\end{figure*}

\section{The Galaxy catalog }
\label{sec3}

The eleventh data release used in this analysis is part of the Baryon Oscillation Spectroscopy Survey (BOSS) experiment  and  comprises the penultimate release from the SDSS third phase (SDSS-III) in a volume of 13 Gpc$^3$. As the other SDSS-III  data releases, the DR11 was divided into two samples: LOWZ (3 Gpc$^3$) and CMASS (10 Gpc$^3$). The LOWZ mapped red galaxies tripling the spatial density provided by the previous SDSS-II for low-redshift galaxy population in range $ 0.15<z<0.43$ whereas the CMASS selection algorithm targets higher redshift galaxies in range $0.43 < z < 0.7$ \cite{Anderson14}.  The CMASS sample contains 543,116 LRGs, i.e., 133,779 LRGs more than the tenth data release. The high DR11 galaxy density allows to compute the 2PACF in five narrow redshift shells ($\delta z = 0.01$ and 0.02) with redshift mean  $\bar{z} = 0.57, 0.59, 0.61, 0.63, 0.65$, as shown in Table \ref{table1}. It is worth observing that we consider  shells that are separated by a redshift interval of 0.005 to avoid correlations between them.

\section{ BAO measurement}
\label{sec4}

As extensively discussed in Ref.~\cite{Carvalho} (see also \cite{Alcaniz,deCarvalho}), the application of the 2PACF to the data usually exhibits more than a single bump which, in general, is due to systematic effects present in the galaxy catalogs. Previously in the literature (see, e.g., \cite{Carnero}), the usual procedure to identify which one corresponds to the real BAO scale was to compare the bump scales observed in the 2PACF with the prediction of the standard cosmology obtained from Eqs. (\ref{expected}) and (\ref{xi_e}).  Here, in order to perform an analysis as model-independent as possible, we adopt the two criteria to identify the real BAO scale introduced in Sec. \ref{sec4} of Ref.~\cite{Carvalho}, namely, the ``bin size'' and ``shifts of the galaxies angular coordinates'' criteria.

\begin{figure} [h!]
\centering
\hspace{-0.8cm}\includegraphics[width=8.5cm, height=6.0cm]{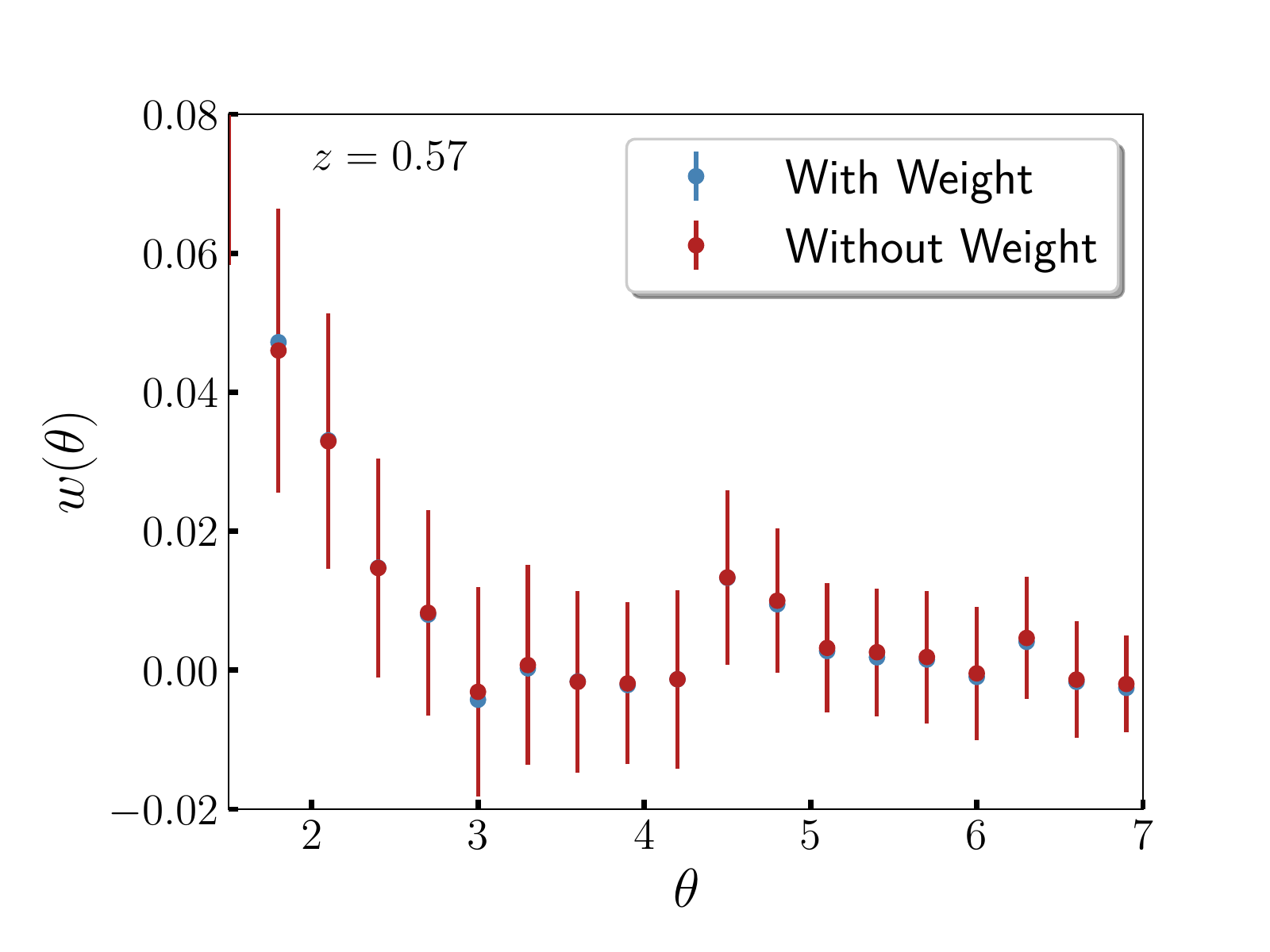}
\caption{The comparison of our 2PACF analyses with and without considering total galaxy weight parameter.}
\label{comparison}
\end{figure}

The BAO signature in our analyzes are fully robust under systematic effects (as observed in Fig. (\ref{comparison})) such as the total galaxy weight parameter, which combines the angular systematics weight with the fibre collision and redshift failure nearest-neighbour weights, producing a quantitative total weight of these effects.

After identifying the true BAO bumps, we obtain the angular BAO scale using the method of Ref.~\cite{Sanchez11}, which {parameterises} the 2PACF as a sum of a power law and a Gaussian peak, i.e.,
\begin{equation} 
w_{FIT}(\theta) = A  + B \theta^\nu + C e^{-\frac{(\theta-\theta_{FIT})^2}{2 \sigma_{FIT}^2}} \, ,
\label{eq:fit}
\end{equation}
where $A,B,C,\nu$, and $\sigma_{FIT}$ are free parameters, and $\theta_{FIT}$ and $\sigma_{FIT}$ correspond, respectively, to the position of the acoustic scale  and the width of the bump. As physically expected, the results of figure \ref{fig1} show a clear change in the $\theta_{FIT}$ toward smaller values as $z$ increases.

\begin{table}
\centering
 \begin{tabular}{|c|c|c|c|c|c|}
\hline
$\Omega_bh^2$& $\Omega_ch^2$ &   100$\Theta$ & $\tau$ & ${\cal{A}}_se^9$ &  $n_s$\\
\hline
\hline
0.0226&  0.112& 1.04 & 0.09  & 2.2  & 0.96\\
\hline 
\end{tabular}
\caption{Values of the cosmological parameters assumed in the calculation of the correction factor $\alpha$.}
\label{table2}
\end{table}

\begin{table}
\centering
\begin{tabular}{ccccc}
\hline
$\bar{z}$&  \,\,\,$\alpha(z, \delta z) [\%]$ & \,\,$\theta_{FIT} [{\rm deg}]$  & 
\,\,\,$\sigma_{\theta_{BAO}}  [{\rm deg}]$ & \,\,\,$\theta_{BAO} [{\rm deg}]$  \\
\hline
\,\,0.57 &  0.28  & 4.61 & 0.40 & 4.62  \\
\,\,0.59 &  0.32  & 4.36 & 0.35 & 4.37  \\
\,\,0.61 &  0.41  & 3.85 & 0.33 & 3.86  \\
\,\,0.63 &  0.56  & 3.86 & 0.42 & 3.88  \\
\,\,0.65 &  1.44  & 3.49 & 0.17 & 3.54  \\
\hline
\end{tabular}
\caption{Values of the cosmological parameters assumed in the calculation of the correction factor $\alpha$.}
\label{table3}
\end{table}

\begin{figure} 
\centering
\hspace{-0.8cm}\includegraphics[width=9.8cm, height=7.4cm]{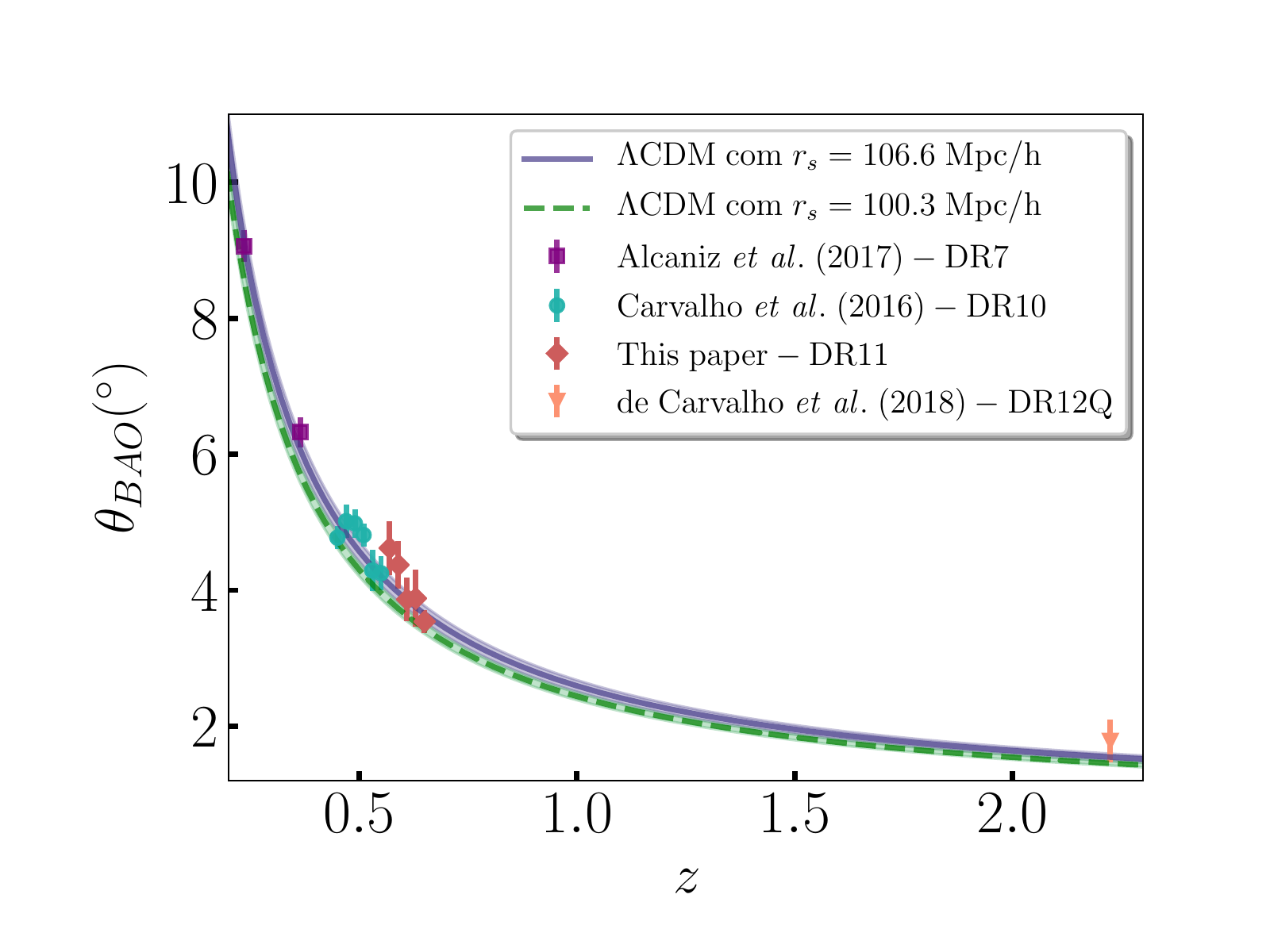}
\caption{Angular BAO scale, $\theta_{BAO}$, as a function of the redshift, $z$, with 13 data points 
obtained using a model-independent procedure through the analyses of the 2PACF using galaxy 
the catalogs from SDSS DR7, DR10, and DR11, also the quasar catalog from DR12. The curves correspond to the $\Lambda$CDM prediction with the sound horizon fixed at the WMAP-9yr ($r_s=106.6$ Mpc/h) and Planck  ($r_s=100.3$ Mpc/h) values with $r_s$ error as a shadow.}
\label{theta-vs-z}
\end{figure}

\begin{figure*}
\centering
 \includegraphics[scale=0.32]{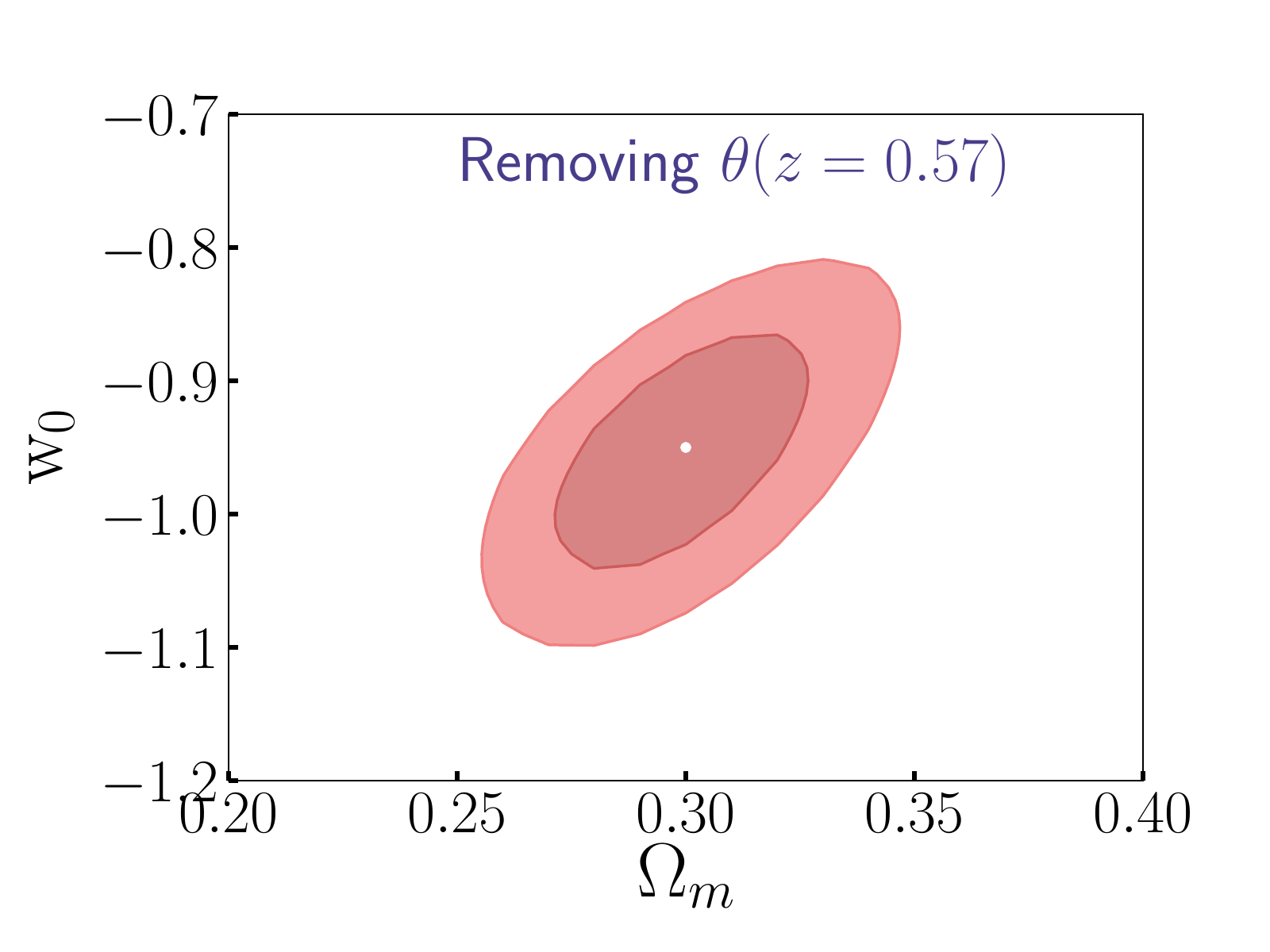}
 \includegraphics[scale=0.32]{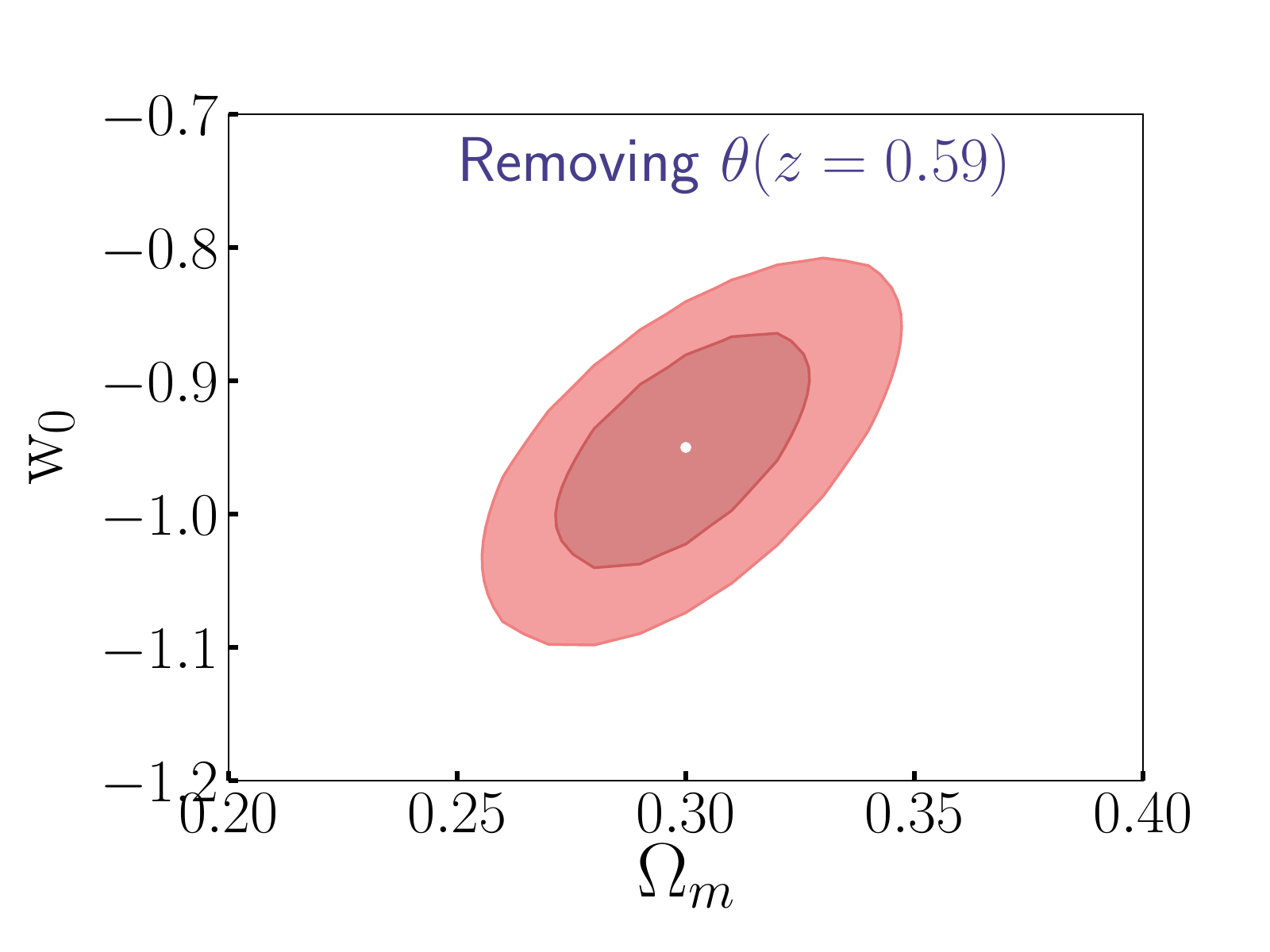}
 \includegraphics[scale=0.32]{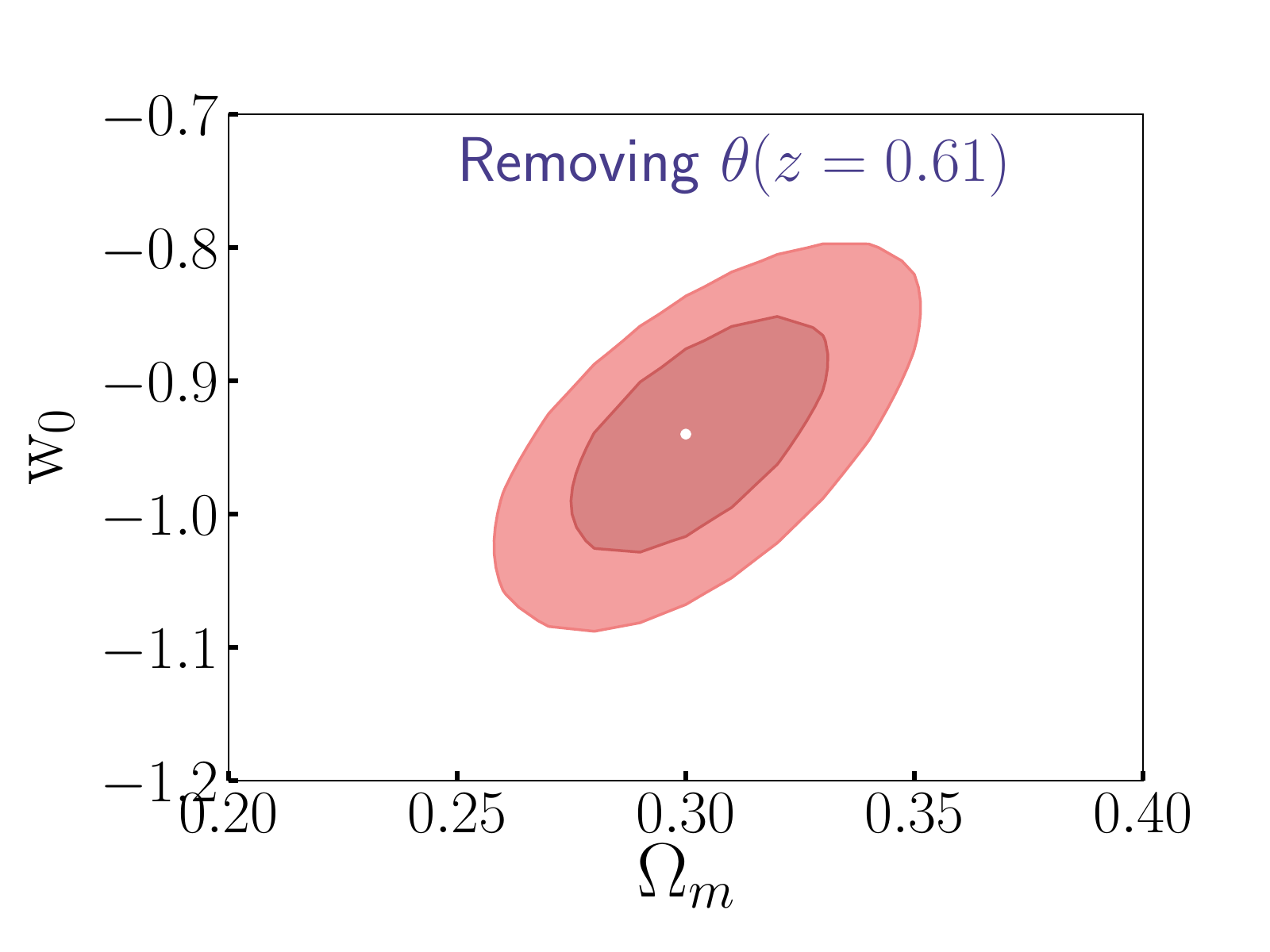}
 \includegraphics[scale=0.32]{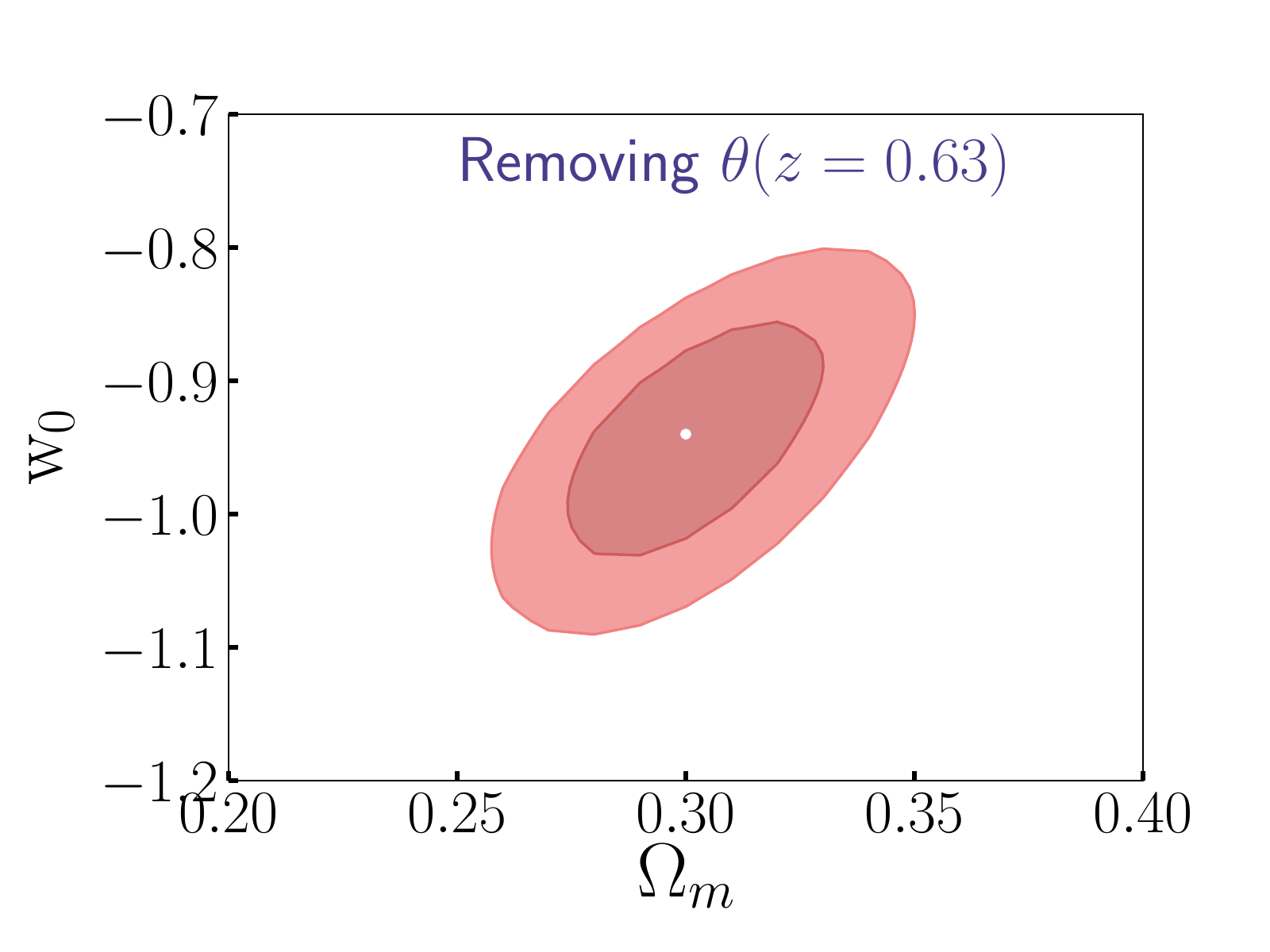}
  \includegraphics[scale=0.32]{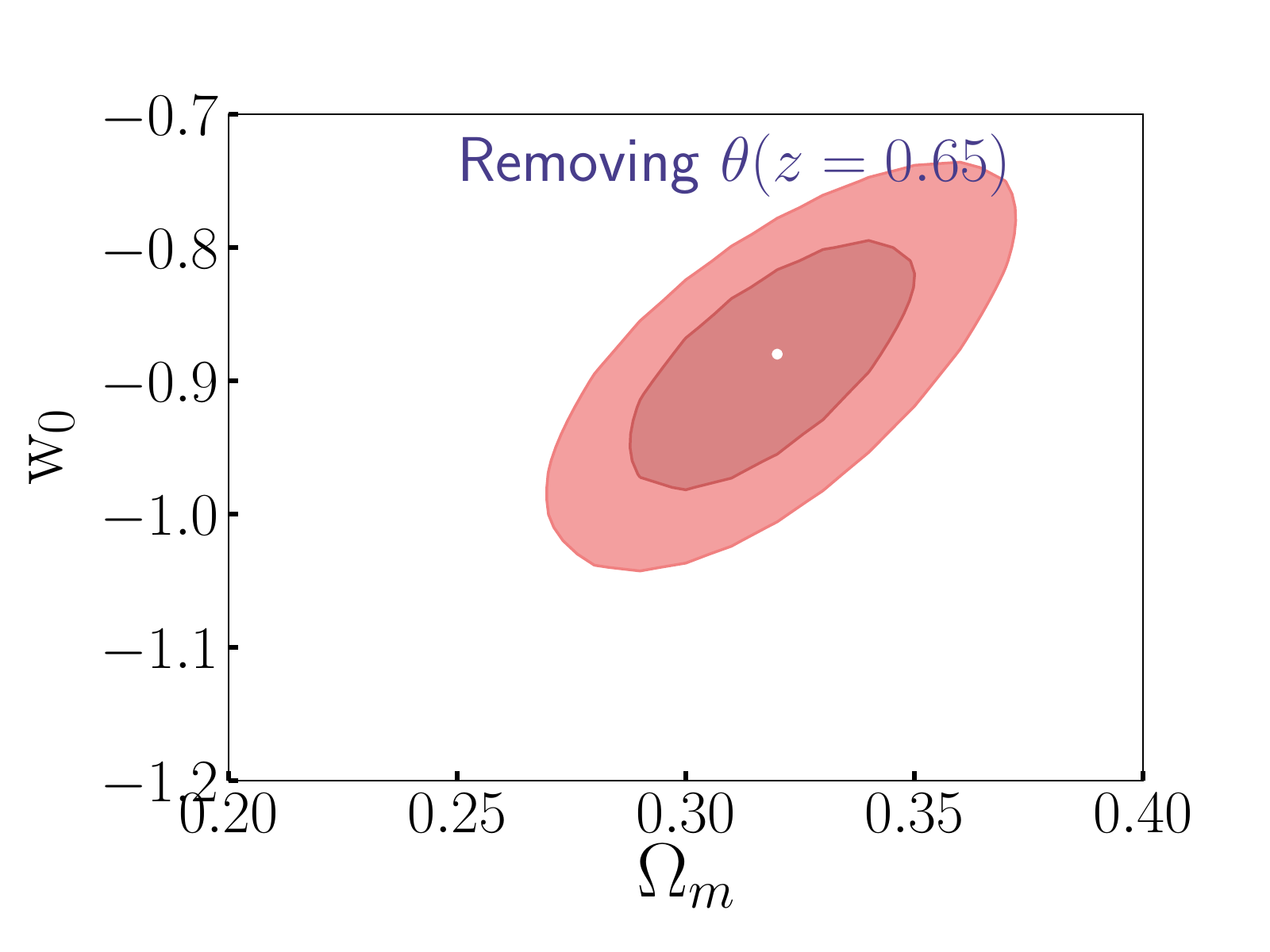}
\caption{The contours removing one data of the $\bar{z} = 0.57, 0.59, 0.61, 0.63, 0.65$ for each time.}
\label{remdata}
\end{figure*}

\begin{figure*} 
\begin{center}
\includegraphics[scale=0.44]
{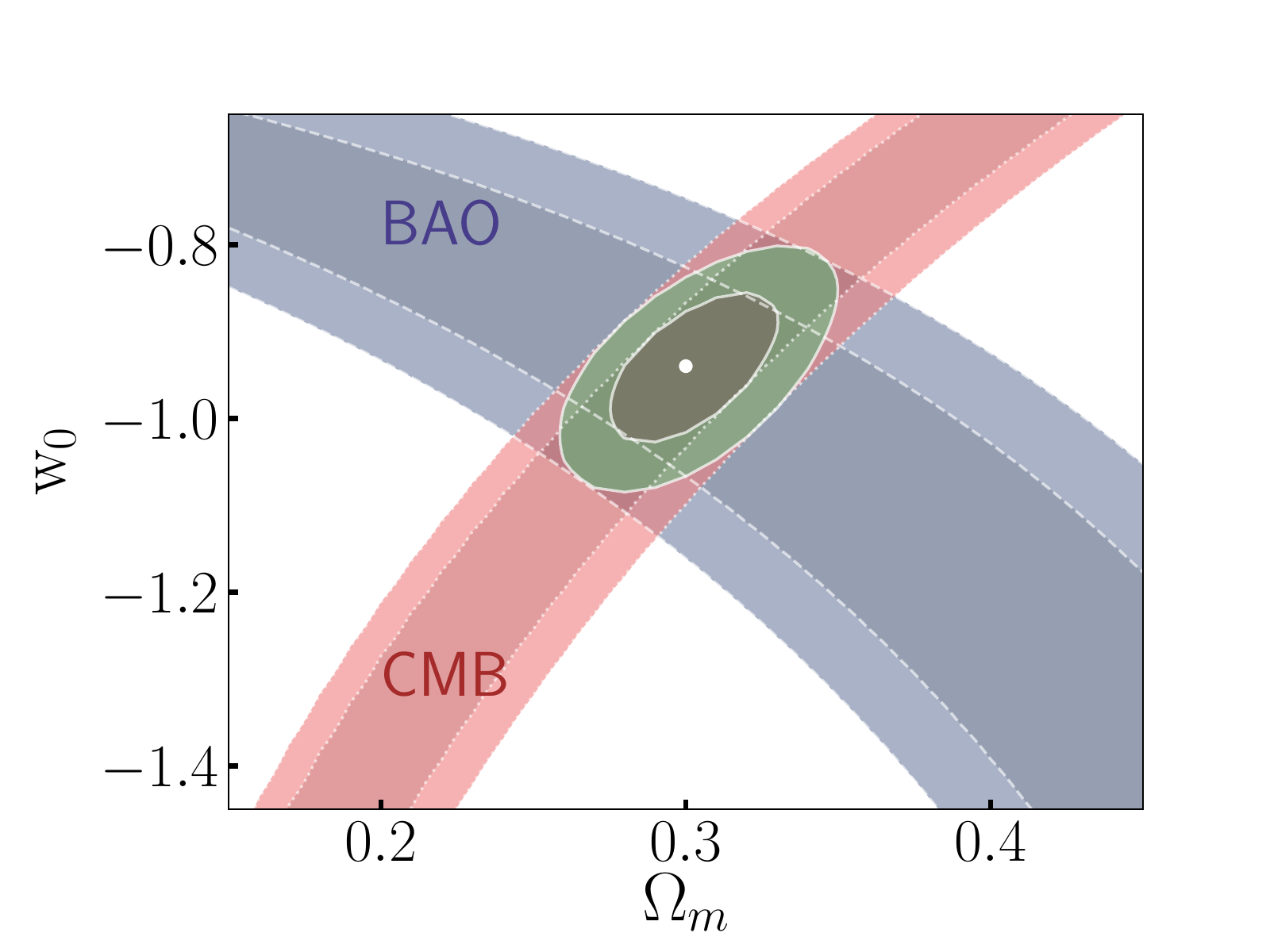}
\hspace{0.4cm}
\includegraphics[scale=0.44]
{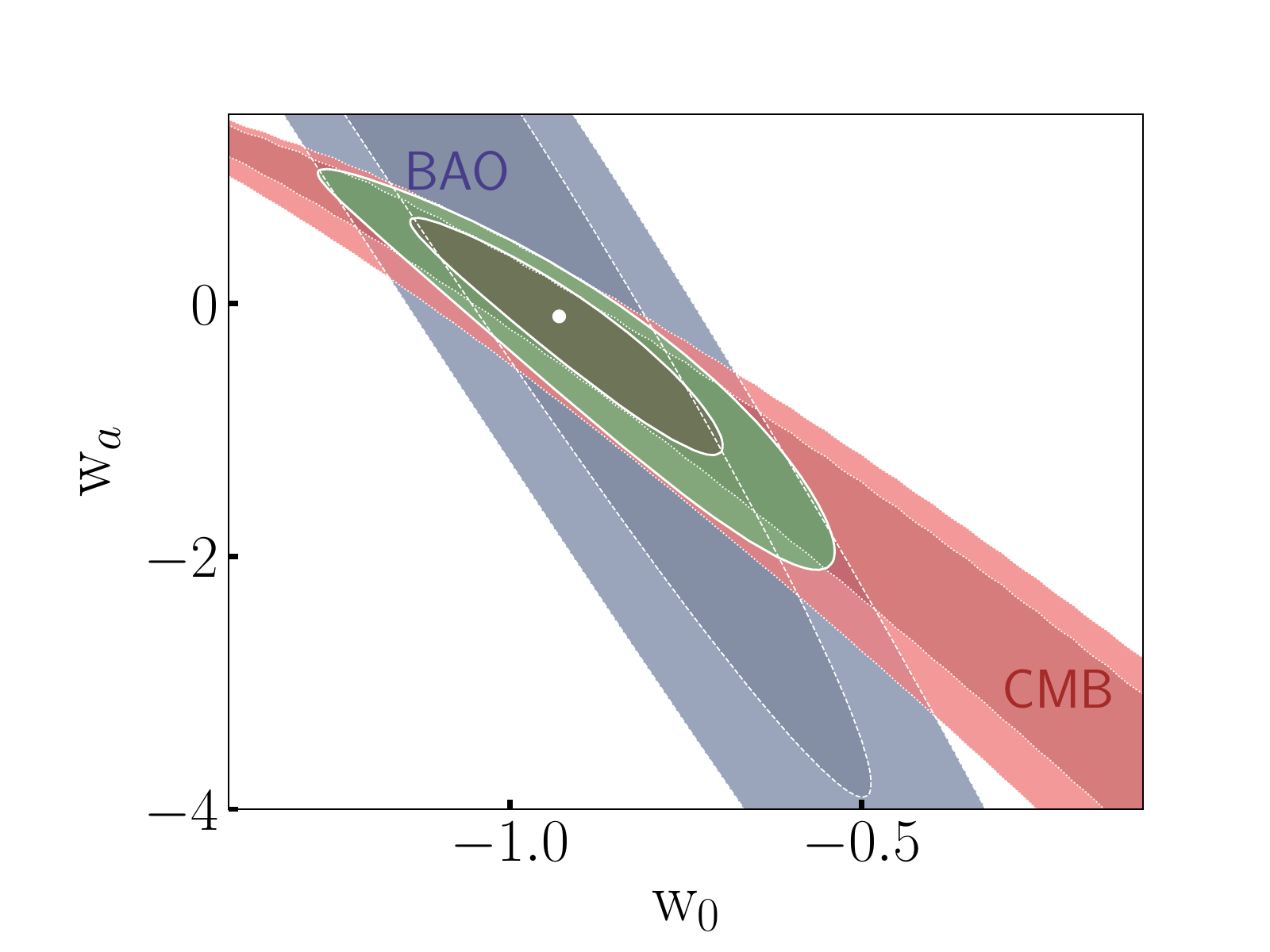}
\end{center}
\caption{{\it{Left)}} The $\Omega_m$ - w$_0$ plane obtained from the $\theta(z)$ data displayed in Table \ref{table3}. Note that the combination between $\theta_{\rm{BAO}}(z)$ and CMB sharply limits the allowed range of the cosmological parameters. {\it{Right)}} The same as in the previous panel for the w$_0$ - w$_a$ plane.
}
\label{contours}
\end{figure*}

As discussed in Refs.~\cite{Sanchez11,Carvalho,Alcaniz,deCarvalho}, the true BAO scale, $\theta_{BAO}$, and the values of $\theta_{FIT}$ obtained from Eq. (\ref{eq:fit}) do not coincide since $\delta z \neq 0$. In order to account for the projection effects due to the width of the redshift shell, we calculate the angular correlation function given by Eqs. (\ref{expected}) and (\ref{xi_e}) for $\delta z = 0$ and $\delta z = 0.01$ and 0.02 (according to Table \ref{table1}). The correction factor $\alpha$ is obtained by comparing the peak position in both cases, i.e.,
\begin{equation}
 \alpha = \frac{\theta_E^0 - \theta_E^{\delta z}}{\theta_E^0}\;.
\end{equation}
To perform this calculation we  have to assume a fiducial cosmology\footnote{Although $\alpha$ very small, this is the first source of model-dependence in our analysis.}. Here we assume a flat $\Lambda$CDM cosmology with the values of the baryon ($\Omega_bh^2$) and cold dark matter ($\Omega_ch^2$) densities, the ratio between the sound horizon and the angular diameter distance at decoupling ($\Theta$),  the optical depth to reionization ($\tau$), the overall normalization of the primordial power spectrum (${\cal{A}}_s$), and the the effective tilt ($n_s$) given in Table \ref{table2}. We also consider purely adiabatic initial conditions and set sum of neutrino masses to 0.06 eV.  

After corrections, the values of the BAO angular scale are given by 
\begin{equation}
 \theta_{BAO}(z, \delta z) = \theta_{FIT} + \alpha(z, \delta z)\theta_{E}^{0}(z)\;,
\end{equation}
and displayed in Table \ref{table3}. As expected, we note that the largest correction occurs for the shell with  $\delta z = 0.02$ ($\Delta z = [0.64, 0.66]$), i.e., $\alpha = 1.44\%$. It is worth mentioning that changing the fiducial cosmology (e.g., considering a time-dependent dark energy equation-of-state) does not alter significantly these results, as shown in Ref.~\cite{Carvalho}. Therefore, given the small width of the shells considered in our analysis, it is possible to say that our measurements of $\theta_{BAO}$ are almost cosmological model-independent\footnote{For large shells of redshift, e.g. $\delta z \simeq 0.1$, the $\alpha$ correction increases significantly and  becomes strongly dependent on the fiducial cosmology adopted in the analysis (see Fig. (3) of Ref.~\cite{Carnero}).}.

We also performed tests to validate how rare (or outlier) is each of the 5 measured values  of $D_A(z)/r_s$. Our analysis determined their individual impact by removing one of them each time, so that we performed 5 tests with a set of 13-1=12 values each time. Then we quantify how different are these results by comparing the relative difference 
with respect to the result considering the 8+5=13 measurements. The results, displayed in the Fig (\ref{remdata}) and Table \ref{table8}, show that all 5 measurements are well-behaved.

\begin{table}
\centering
\begin{tabular}{| c | c | c |}
\hline
data removed  & $\Omega_M$ & $w_0$ \\
\hline
$z=0.57$ & 0.3 & -0.95 \\
\hline
$z=0.59$ & 0.3 & -0.95 \\
\hline
$z=0.61$ & 0.3 & -0.94 \\
\hline
$z=0.63$ & 0.3 & -0.94 \\
\hline
$z=0.65$ & 0.32 & -0.88 \\
\hline
\end{tabular}
\caption{The best-fit removing one data obtained.}
\label{table8}
\end{table}

\subsection{\label{sec:level2.1} The Covariance Matrix}

In order to compute the 2PACF error bars, we follow \cite{deCarvalho} that showed that the jackknife method provides a better result in the acoustic scale determination when compared with other methods, such as bootstrap.

The jackknife method consists of computing $N$ times the correlation function  by removing part of the real sample in each computation.  In order to avoid correlations between the different sub-samples we split the real data into 30 pieces. Then we assume $N$ sub-samples with a $(N-1)/N$ fraction of the real volume.  The covariance matrix for the jackknife method assuming $\{w_1,w_2, \cdots, w_N\}$ sub-samples is given by

\begin{equation}
C(w_i,w_j)=\frac{(N-1)}{N} \sum_{n=1}^{N}(w_n(\theta_i)-\bar{w}(\theta_i)) \; (w_n(\theta_j)-\bar{w}(\theta_j))\;,
\end{equation}
where
\begin{equation}
\bar{w}(\theta_i)=\sum_{n=1}^{N}\frac{w_n(\theta_i)}{N}\;.
\end{equation}
The 2PACF error bar for each $\theta$ bin corresponds to the square root of its respective diagonal element of the covariance matrix.

\section{Cosmological Constraints} 
\label{sec5}

In terms of the angular diameter distance, $D_A = r(z)/(1+z)$, where $r(z)$ is given by Eq. (\ref{rz}), the BAO angular scale can be written as
\begin{equation} \label{tz}
 \theta_{BAO} = \frac{r_s}{(1 + z) D_A(z)}\;,
\end{equation}
where the sound horizon is defined as $r_s = \int_{z_d}^{\infty}{dz}{c_s(z)/H(z)}$ with $z_d$ being the redshift of the drag epoch and $c_s(z)$ the sound speed of the photon-baryon fluid. The evolution of $\theta_{BAO}$ with redshift is shown in Fig.~(\ref{theta-vs-z}) along with the data points obtained in this analysis (gray) and also in Refs.~\cite{Carvalho} (red),~\cite{Alcaniz} (green), and \cite{deCarvalho} (orange), which used the same methodology adopted here. The thick gray and green curves correspond, respectively, to the prediction of the standard cosmology assuming $\Omega_m = 0.27$ and two different values of  sound horizon, i.e., $r_s = 106.61 \pm 3.47$ Mpc/h (68.3\% C.L.), obtained by the Wilkinson Microwave Anisotropy Probe (WMAP) collaboration~\cite{wmap9} and $r_s = 100.29 \pm 2.26$ Mpc/h (68.3\% C.L.), derived by the Planck Collaboration~\cite{planck}.

In order to derive cosmological constraints from the $\theta_{BAO}$ data displayed in Table \ref{table3}, we consider two different cosmologies, namely, a varying dark energy model with equation-of-state (EoS) parameter given by $w(a) = w_0 + w_a(1-a)$ ($w_a$CDM) and a scenario with a constant EoS parameter, $w_a = 0$ and $w_0 \neq -1$ ($w$CDM), which includes the standard $\Lambda$CDM cosmology ($w_0 = -1$) as a particular case. Assuming the WMAP-9yr estimate of the sound horizon\footnote{This is the second source of model-dependence in our analysis since the most precise measurements of $r_s$ are obtained using CMB data assuming a given cosmology.}, we show the resulting cosmological constraints for the combined DR7, DR10, DR11, and DR12Q $\theta_{BAO}$ data in Fig.~(\ref{contours}). Clearly, this combination of data alone is consistent with a wide range of $w_0$ and $w_a$ values (gray contours). On the other hand, tighter constraints are obtained by combining our $\theta_{BAO}$ measurements with CMB data, e.g., with the CMB shift parameter, ${\cal{R}} = \sqrt{\Omega_m}\int_{0}^{z_{ls}}dz {H_0/H(z)}$, where $z_{ls}$ is the redshift of the last scattering surface. In order to avoid double counting of information with the sound horizon from WMAP Collaboration used in the $\theta_{BAO}$ analysis, we use ${\cal{R}} = 1.7407 \pm 0.0094$, as  given the Planck Collaboration~\cite{planck}. The joint analysis provides: $(w_0, w_a) = (-0.93 \pm 0.145, -0.1 \pm 0.605)$ and $(\Omega_m, w_0) = (0.30 \pm 0.01, -0.94 \pm 0.055)$ at 68.3\% (C.L.).

\begin{figure} 
\begin{center}
\includegraphics[width=7.8cm, height=5.9cm]{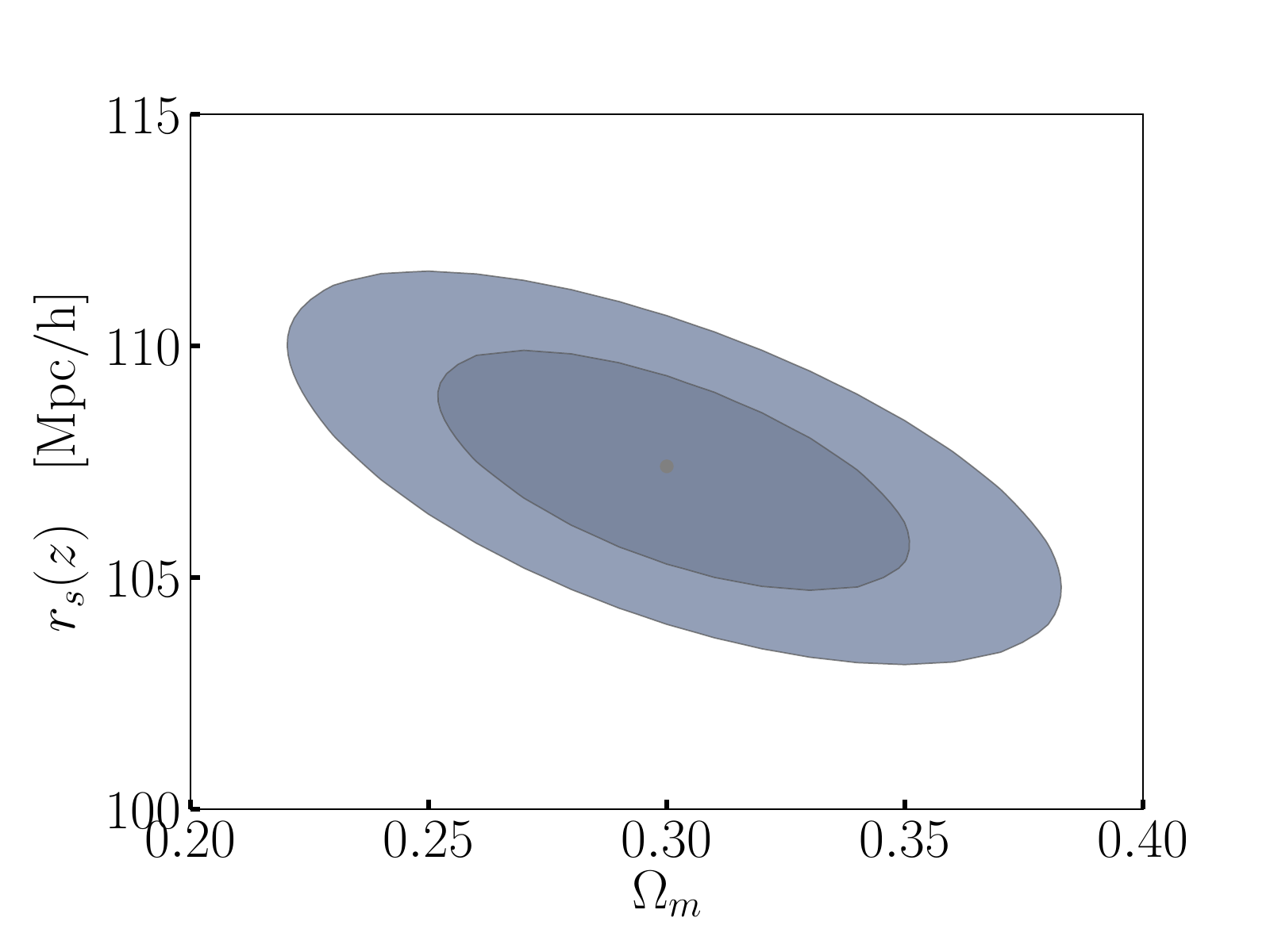}
\end{center}
\caption{The confidence contours in the $\Omega_m$ - $r_s$ plane for $\Lambda$CDM model assuming the estimate of the matter density parameter from type Ia supernova data}
\label{rsomega}
\end{figure}

As mentioned earlier, from Eq.~(\ref{tz}), an independent estimate of the sound horizon directly from the distribution of galaxy can be directly obtained assuming a given cosmology. In order to compare our results with the CMB estimates of $r_s$ from WMAP and Planck collaborations, we also assume a flat $\Lambda$CDM cosmology. We find $r_s = 104.0 \pm 4.0$ Mpc/h (68.3\% C.L.), which is in good agreement with both WMAP-9yr and Planck estimates. Considering a prior on the matter density parameter, $\Omega_m = 0.295 \pm 0.034$ at 68.3\% (C.L.), as given by current supernova data~\cite{sne}, we obtain our final estimate of the sound horizon, i.e., $r_s = 107.4 \pm 1.7 h^{-1}$ Mpc at 68.3\% C.L. (Fig. (\ref{contours})), which is  in good agreement with WMAP-9yr results and $\sim 3\sigma$ off from the central value obtained by the Planck collaboration. For completeness, we summarize the current measurements and estimates of the sound horizon in Table \ref{table5}.

\begin{table}
\centering
\begin{tabular}{ cc }
 \hline  
 Reference &  $r_s$ (Mpc/h) \\
 \hline
 \,\, WMAP-9yr   \cite{wmap9}            & $106.61 \pm 3.47 $ \\
  \,\, Planck \cite{planck}             & $100.29  \pm 2.26 $ \\
  \,\,  Heavens \textit{et. al.} (2014) \cite{Heavens} & $ 101.9 \pm 1.9$ \\
  \,\, Carvalho \textit{et. al.} (2016) \cite{Carvalho} & $ 107.6 \pm 2.3$ \\
  \,\, Verde \textit{et. al.} (2017) \cite{Verde17} & $102.3 \pm 1.6$ \\
  \,\,  This paper  & $107.4 \pm 1.7$ \\
  \hline
 \end{tabular}
 \caption{Current estimates of the sound horizon $r_s$.}
 \label{table5}
\end{table}

\section{Conclusions} 
\label{sec6}

In this paper we have extended previous analyses \cite{Carvalho,Alcaniz} and reported five new measurements of the angular BAO scale, $\theta_{BAO}$, in the redshift interval $0.43 < z < 0.70$ using the LRGs sample of the SDSS DR11. Differently from  the 2PCF analysis usually adopted in the literature, which assumes a fiducial cosmology in order  to transform the measured angular positions and redshifts into distances, the 2PACF methodology adopted here involves only the angular separation between pairs, providing a measurement of $\theta_{BAO}$ almost model-independently. 

Two sources of model-dependence, however, can be directly identified in our analysis, namely, the theoretical model assumed in the calculation of the $\alpha$-correction discussed in Sec. \ref{sec4} and the value of $r_s$ adopted in the statistical analysis presented in Sec. \ref{sec5}. The former introduces a 1-2\% correction in the $\theta_{BAO}$ position (see Table \ref{table3}), given the width of the redshift shell considered in our analysis ($\delta z = 0.1 -0.2$), whereas the latter depends basically on the physics of the early universe, which is similar in most of the viable cosmological models.

The five data points reported in this work, together with the eight data points of Refs.~\cite{Carvalho,Alcaniz}, span the redshift interval $z \in [0.2,0.66]$ for LRGs data and improve our previous constraints on the dark energy parameters $w_0$ and $w_a$, as shown in Sec. 5. When combined with CMB data, we have found $(w_0, w_a) = (-0.93 \pm 0.145, -0.1 \pm 0.605)$ and $(\Omega_m, w_0) = (0.30 \pm 0.01, -0.94 \pm 0.055)$ at 68.3\% (C.L.), which is in agreement with the standard $\Lambda$CDM model as well as some of its extensions. 

Finally, given the current disagreement in what concerns the value of the $r_s$ estimated by the two most recent CMB data sets, an important task nowadays is to derive independent estimates of this quantity from other observables.  Here, we have obtained a new measurement from low-$z$  galaxy clustering data, $r_s =  107.4 \pm 1.6 $ Mpc/h (68.3\% C. L.). Our result is in good agreement with the one derived by the WMAP Collaboration but in some tension with the current estimate from the Planck Collaboration.

\acknowledgments

The authors thank CNPq, CAPES and FAPERJ for the grants under which this 
work was carried out.  J. C. Carvalho is also supported by the DTI-PCI  program of the Brazilian Ministry of Science, Technology, and Innovation (MCTI).


\end{document}